\documentclass[12pt]{article}
\usepackage{amsfonts}
\usepackage{amsmath,amssymb}
\usepackage[utf8]{inputenc}
\usepackage[left=2cm,right=1.5cm,top=1.5cm,bottom=1.5cm,bindingoffset=0cm]{geometry}

\numberwithin{equation}{section}
\input{tcilatex}
\begin{document}

\title{Poisson structures of some heavenly type dynamical systems}
\author{Y.A. Prykarpatskyy \\ 
	Dept. of Applied Mathematics, University of Agriculture in Krakow, \\
	Al. Mickiewicza 21, 31-120,
	Kraków, Poland}
\maketitle

\begin{abstract}
The paper investigates the Poisson structures associated with dynamical systems of the heavenly type, focusing on the Mikhalev-Pavlov
and Plebański equation. The dynamical system is represented as a Hamiltonian system on a functional manifold, and Poisson brackets are defined based on a non-degenerate Poisson operator. The study explores the Lax-type integrability and bi-Hamiltonian properties of the systems, revealing the existence of compatible Poisson operators. The Lie-algebraic approach, particularly the AKS-algebraic and $\mathcal{R}$-structure schemes, is employed to analyze the holomorphic loop Lie algebra, providing insights into the Lie-algebraic structure of heavenly equations. The Mikhalev-Pavlov and Plebański equations are studied in detail, and the associated Poisson brackets for specific coordinate functions are derived, revealing interesting mathematical properties. The paper establishes a foundation for understanding the symplectic structures associated with heavenly-type dynamical systems.
\end{abstract}


\section{Introduction}

Let us consider a dynamical system 
\begin{equation}  \label{int01}
du/dt=K[u],
\end{equation}
on the functional manifold $M$, where $u=(u_1(x),...,u_m(x))^\top \in M\subset \mathbb{C}_{2\pi}^\infty(\mathbb{R},%
\mathbb{R}^m)$, and $K:M\to T(M)$ is a vector field, $T(M)$ is a vector tangent
space to the infinite-dimension $2\pi$-periodic manifold $M$. For any two
functionals $F_1,F_2\in \mathcal{D}(M)$ one can define the Poisson brackets 
\begin{equation}  \label{int02}
\{ F_1,F_2\}_\theta (u)=\left(\left<\mathrm{grad}\ F_1\mid \theta\mathrm{grad}\ F_2\right>\right),
\end{equation}
where $\theta:T^*(M)\to T(M)$ is a non-degenerated Poisson and
skew-symmetrical operator, $\mathrm{grad}\ F=F^{\prime \star}_u[u]\cdot \mathbf{1%
}$, $u\in M$, $"^{\prime }"$ is the Fr\'echet derivative. The scalar product
is given as 
\begin{equation}  \label{int03}
\left(\left< a \mid b \right>\right)=\int\limits_{\mathbb{T}^m}\left<a\mid b\right>dx,
\end{equation}
$a,b\in \mathbb{C}_{2\pi}^\infty(\mathbb{R},\mathbb{R}^m)$, where $%
\left<\cdot \mid \cdot\right>$ denotes the Cartesian product in $\mathbb{R}^m$. Dynamical system %
\eqref{int01} is called Hamiltonian if it can be rewritten as 
\begin{equation}  \label{int04}
du/dt=\{H,u \}_\theta=-\theta\mathrm{grad}\ H.
\end{equation}
where $\theta$ is a matrix skew-symmetrical integro-differential operator
which satisfies the Jacoby identity. 
The coordinate components $u_i:M\to\mathbb{R}$, $i=\overline{1,m}$ are
considered as local functionals on $M$,  being represented as
\begin{equation}  \label{int05}
u_i(x)=\int\limits_{0}^{2\pi}u_i(y)\delta(x-y)dy.
\end{equation}
Then the Poisson brackets for coordinate functions $u_i,u_j\in M\to \mathbb{R}$, $i,j\in\overline{1,m}$ are given by
\begin{equation}  \label{int06}
\{u_i(x),u_j(y)\}_\theta=\int\limits_{0}^{2\pi}\delta(x-z)%
\theta_{ij}[u_1(z),...,u_m(z)]\delta(y-z)dz=\theta_{ij}[u_1(x),...,u_m(x)]%
\delta(y-x),
\end{equation}
where we used the expression
\begin{equation}
\mathrm{grad}\ u_i(x)=(0,...,\delta(x-y),...,0).
\end{equation}
If the dynamical system \eqref{int01} is Lax-type integrable and bi-Hamiltonian then there exist pair of
compatible Poisson operators $\theta$ and $\eta$
that 
\begin{equation}  \label{int07}
du/dt=-\vartheta\mathrm{grad}\ H=-\eta\mathrm{grad}\ \tilde{H},
\end{equation}
where $\gamma_0=H$ and $\gamma_1=\tilde{H}$ are the corresponding local conservation laws of
the system \eqref{int07}. The operators $\vartheta,\eta:\mathbb{T}^*(M)\to\mathbb{T}(M)$ allow to define an
infinite set of the conservation laws $\gamma_k:M\to\mathbb{R}$, $k\in\mathbb{Z}$, 
and respectively infinite set of the non-degenerate
skew-symmetric and Poisson operators for \eqref{int07} 
\begin{eqnarray}  \label{int08}
&&\mathrm{grad}\ \gamma_k=(\vartheta^{-1}\eta)^k\mathrm{grad}\ \gamma_0,
\end{eqnarray}
where $\vartheta^{(k)}=\vartheta(\vartheta^{-1}\eta)^k$  and $k\in\mathbb{Z}$. Then the system \eqref{int01} can be rewritten in
the following form 
\begin{equation}  \label{int09}
\partial l/\partial t=\partial p/\partial x +[p,l],
\end{equation}
where $l=l[u;\lambda]$, $p=p[u;\lambda]$ are the linear operators which act
on the some functional space, $u\in M$, $\lambda\in\mathbb{C}^1$ is a
spectral parameter. The operator $l[u;\lambda]$ is called the Lax operator
for the system \eqref{int01}. The Lax operator can be an element of some
linear space 
which can by analyzed by means of the Lie-algebraic approach developed in 
\cite{FaTa}.

Recently in series of papers \cite{HePrBlPr,HePr-1,PrHePr} the AKS-algebraic
and related $\mathcal{R}$-structure schemes \cite{FaTa,ReSe} were applied to
the holomorphic loop Lie algebra $\mathcal{\tilde{G}}:=\widetilde{diff}(%
\mathbb{T}^{m})$ of vector fields on torus $\mathbb{T}^{m}$, $m\in\mathbb{N}$. 
It allows to study the orbits of the corresponding coadjoint actions
on $\mathcal{\tilde{G}}^{\ast}$, closely related to the classical
Lie--Poisson type structures. It was shown that heavenly type equations are the
result of the compatibility conditions of the two constructed commuting
flows on the coadjoint space $\mathcal{\tilde{G}}^{\ast}$, generated by a
chosen seed element $\tilde{l}\in\mathcal{\tilde{G}}^{\ast}$ and some
Casimir invariants. To proceed further we need to recall the necessary terms
from the developed recently Lie-algebraic approach to the theory of heavenly
equations.

\section{Lie-algebraic structure of heavenly type equations}

Let $\tilde{G}_{\pm}:=\widetilde{Diff}_{\pm}(\mathbb{T}^{n}),$ $n\in \mathbb{N}$, 
be subgroups of the loop diffeomorphisms group $\widetilde {Diff}(%
\mathbb{T}^{n}):=$ $\{\mathbb{C}\supset\mathbb{D}^{1}\rightarrow Diff(%
\mathbb{T}^{n})\}$, holomorphically extended in the interior $\mathbb{D}%
_{+}^{1}\subset\mathbb{C}$ and in the exterior $\mathbb{S}_{-}^{1}\subset%
\mathbb{C}$ \ regions of the unit circle $\mathbb{S}^{1}\subset\mathbb{C}%
^{1},$ such that for any $g(\lambda)\in\tilde{G}_{\pm},$ $\lambda\in\mathbb{%
\ }\mathbb{D}_{-}^{1},$ $g(\infty)=1\in Diff(\mathbb{T}^{n}).$ The
corresponding Lie subalgebras $\mathcal{\tilde{G}}_{\pm }:=\widetilde{diff}%
_{\pm}(\mathbb{T}^{n})$ of the loop subgroups $\ \tilde {G}_{\pm}$ are
vector fields on $\mathbb{T}^{n}$ holomorphic, respectively, on $\mathbb{D}%
_{\pm}^{1}\subset\mathbb{C}^{1},$ where one can \ single out two cases: the
first one, if for any $\tilde{a}(\lambda)\in\mathcal{\tilde{G}}_{+}$ the
value $\tilde{a}(0)=0,$ and the second one, if for any $\tilde{a}(\lambda)\in%
\mathcal{\tilde{G}}_{-}$ the value $\tilde{a}(\infty)=0.$ The natural loop Lie
algebra splitting $\mathcal{\tilde{G}=\tilde{G}}_{+}+$ $\mathcal{\tilde{G}}_{-}$ can
be identified with a dense subspace of the dual space 
$\mathcal{\tilde{G}}^{\ast}$ through the pairing 
\begin{equation}
(\tilde{l}\mid \tilde{a}):=\underset{\lambda\in\mathbb{C}}{res}(\lambda
^{-p}l(x,\lambda)\mid a(x,\lambda))_{H^{q}},  \label{eq1.1}
\end{equation}
for some fixed $p\in\mathbb{Z}$ and $q\in\mathbb{Z}_{+}$. We take a loop
vector field $\tilde{a}\in\Gamma(\tilde{T}(\mathbb{T}^{n}))$ as 
\begin{equation}
\tilde{a}\ =\sum\limits_{j=1}^{n}a^{(j)}(x,\lambda)\frac{\partial}{\partial
x_{j}}:=\left\langle a(x;\lambda)\bigg| \frac{\partial}{\partial x}\right\rangle ,
\label{eq1.2_1}
\end{equation}
and a loop differential 1-form 
$\tilde{l}\in\tilde{\mathcal{G}}^{*}$ 
as 
\begin{equation}  \label{eq1.2_2}
\tilde{l}\ =\sum\limits_{j=1}^{n}l_{j}(x,\lambda)dx_{j}:=\ \left\langle
l(x;\lambda)\mid dx\right\rangle.
\end{equation}
Here $\frac{\partial}{\partial x}$ denoted the gradient operator $\frac{%
\partial}{\partial x}:=\left( \frac{\partial}{\partial x_{1}},\frac{\partial%
}{\partial x_{2}},...,\frac{\partial}{\partial x_{n}}\right) ^{\intercal}$ in the
Euclidean space $\mathbb{E}^{n}$, and $(\cdot,\cdot)_{H^{q}}$ is the Sobolev
type metric on the space $C^{\infty}(\mathbb{T}^{n};\mathbb{R}^{n})\subset
H^{q}(\mathbb{T}^{n};\mathbb{R}^{n})$ for some $q\in \mathbb{Z}_{+}$ 
\begin{equation}
(l(x;\lambda),a(x;\lambda))_{H^{q}}:=\sum\limits_{j=1}^{n}\sum\limits_{|%
\alpha |=0}^{q}\int\limits_{\mathbb{T}^{n}}dx\left( \frac{%
\partial^{|\alpha|} l_{j}(x;\lambda)}{\partial x^{\alpha}}\frac{%
\partial^{|\alpha|}a^{(j)} (x;\lambda)}{\partial x^{\alpha}}\right) ,
\label{eq1.2a}
\end{equation}
where $\partial x^{\alpha}:=\partial x_{1}^{\alpha_{1}}\partial
x_{2}^{\alpha_{2}}...\partial
x_{2}^{\alpha_{n}},|\alpha|=\sum_{j=1}^{n}\alpha_{j}$ for $\alpha\in\mathbb{Z%
}_{+}^{n}$. The Lie commutator of vector fields $\tilde{a},\tilde{b}$ $\in%
\mathcal{\tilde{G}}$ equals 
\begin{align}  \label{eq1.3}
\lbrack\tilde{a},\tilde{b}] =\tilde{a}\tilde{b}-\tilde{b}\tilde {a}%
=\left\langle \left\langle a(x;\lambda)\bigg|\frac{\partial}{\partial x}%
\right\rangle b(x;\lambda)\bigg|\frac{\partial}{\partial x}\right\rangle -\
\left\langle \left\langle b(x;\lambda)\bigg|\frac{\partial}{\partial x}%
\right\rangle a(x;\lambda)\bigg|\frac{\partial}{\partial x}\right\rangle .  \notag
\end{align}
The Lie algebra $\mathcal{\tilde{G}}$ splitting  $\mathcal{\tilde{G}}=\mathcal{\tilde{G}}_{+}\oplus\mathcal{\tilde{G}} _{-}$ 
generates the dual direct sum space 
$\mathcal{\tilde{G}}^*=\mathcal{\tilde{G}}_{+}^*\oplus\mathcal{\tilde{G}} _{-}^*$,
\begin{equation*}
\mathcal{\tilde{G}}_{+}^{\ast}\simeq\lambda^{p-1}\mathcal{\tilde{G}}_{-},\ \
\ \ \ \ \mathcal{\tilde{G}}_{-}^{\ast}\simeq\lambda^{p-1}\mathcal{\tilde{G}}%
_{+},
\end{equation*}
where for any $l(\lambda)\in\mathcal{\tilde{G}}_{-}^{\ast}$ one has assumed the
constraint $\tilde{l}(0)=0$. The projection operator 
$
P_{\pm}\mathcal{\tilde{G}}:=\mathcal{\tilde{G}}_{\pm}\subset\mathcal{\tilde{G}} 
$
allows to introduce a classical $\mathcal{R}$-structure \cite{FaTa,ReSe,Seme}
on the Lie algebra $\mathcal{\tilde{G}}$ as the endomorphism $\mathcal{R}:%
\mathcal{\tilde{G}\rightarrow\tilde{G}},$ where 
\begin{equation}
\mathcal{R}:=\ (P_{+}-P_{-})/2.  \label{eq1.5}
\end{equation}
Then 
related new Lie algebra structure 
\begin{equation}
\lbrack\tilde{a},\tilde{b}]_{\mathcal{R}}:=[\mathcal{R}\tilde{a},\tilde {b}%
]+[\tilde{a},\mathcal{R}\tilde{b}]  \label{eq1.6}
\end{equation}
for any $\tilde{a},\tilde{b}\in\mathcal{\tilde{G}}$ satisfies the standard
Jacobi identity.

Let $\mathrm{D}\mathcal{(\tilde{G}}^{\ast})$ denote the space of smooth
functions on $\mathcal{\tilde{G}}^{\ast}.$ Then for any $f,g\in \mathrm{D}%
\mathcal{(\tilde{G}}^{\ast})$ one can write the canonical \cite%
{BlPrSa,FaTa,ReSe,PrMy} Lie--Poisson bracket 
\begin{equation}
	\{f,g\}:=(\tilde{l},[\nabla f(\tilde{l}),\nabla g(\tilde{l})]),
	\label{eq1.8}
\end{equation}
where $\tilde{l}\in\mathcal{\tilde{G}}^{\ast}$ is a seed element and $\nabla
f,$ $\nabla g\in\mathcal{\tilde{G}}$ \ are the standard functional gradients
at $\tilde{l}\in\mathcal{\tilde{G}}^{\ast}$ with respect to the metric (\ref%
{eq1.1}).

\section{The Mikhalev-Pavlov equation case}

Let us take $\mathcal{\tilde{G}}^{\ast}:=\widetilde{diff}^{\ast}(\mathbb{T}%
^{1})$ and consider the Mikhalev-Pavlov equation \cite{Mikh} 
\begin{equation}  \label{MP01}
u_{xt}+u_{yy}=u_yu_{xx}-u_xu_{xy}
\end{equation}
where $u\in M\subset  C^{\infty}(\mathbb{R}^{2}\times\mathbb{T}^{1})$ and $(t,y;x)\in 
\mathbb{R}^{2}\times\mathbb{T}^{1}$. It was shown in \cite{HePrBlPr} that the
following ''seed''-element $\tilde{l}_0\in\mathcal{\tilde{G}}^{\ast}$ 
\begin{equation}  \label{MP02}
\tilde{l}_0=(\lambda-2u_x)dx
\end{equation}
generates the following commuting flows on $\mathcal{\tilde{G}}^{\ast}$ 
\begin{eqnarray}  \label{MP03}
&&\frac{\partial\psi}{\partial t}+(\lambda^{2}+\lambda u_{x}-u_{y} )\frac{%
\partial\psi}{\partial x}=0 \\
&&\frac{\partial\psi}{\partial y} +(\lambda+u_{x})\frac{\partial\psi}{%
\partial x}=0,  \notag
\end{eqnarray}
satisfied for $\psi\in C^{\infty}(\mathbb{R}^{2}\times\mathbb{T}^{1};\mathbb{%
C}),$ any $(y,t;x)\in\mathbb{R}^{2}\mathbb{\times T}^{1}$ and all $\lambda\in%
\mathbb{C}$. The compatibility condition of the flows \eqref{MP03} gives rise to
the Mikhalev-Pavlov heavenly equation \eqref{MP01}.

We can also take $\tilde{l}_p\in\tilde{\mathcal{G}}^*$
\begin{equation}  \label{MP04}
\tilde{l}_p=\lambda^p(\lambda+u_0)dx
\end{equation}
where we chose functional $u_0(s_1)=-2 u_{s_1}$ as an independent coordinate
with parameter $s_1\in\mathbb{T}^1$.

To find the corresponding Poisson brackets for coordinate $u_0(s_1)$ we remark,
that functional $u_0(s_1)$ has the following representation \cite{HePrBlPr} on 
$\mathcal{\tilde{G}}^*$  at $q=0$: 
\begin{equation}  \label{MP05}
u_0(s_1)=\left( \tilde{l}_p \bigg| \mathrm{grad} (u_0(s_1))\frac{\partial}{\partial x}%
\right) \Rightarrow \left( (\lambda^{p+1}+\lambda^p u_0)dx\bigg| \mathrm{grad} (u_0(s_1)) \frac{%
\partial}{\partial x} \right)= res_\lambda\int\limits_{0}^{2\pi}\lambda^p u_0
\mathrm{grad} (u_0(s_1)) dx,
\end{equation}
from which one can find the expression for the gradient of $u_0(s_1)$: 
\begin{equation}  \label{MP06}
\mathrm{grad} (u_0(s_1))=\lambda^{-(p+1)}\delta(x-s_1)\frac{\partial}{\partial x}
\end{equation}
The latter makes it possible to calculate the Lie-Poisson bracket \eqref{eq1.8} 
\begin{eqnarray}  \label{MP07}
&&\left\{ u_0(s_1),u_0(s_2) \right\}_p=\left(\tilde{l}_n\bigg|\left[\nabla
u_0(s_1), \nabla u_0(s_2)\right]_\mathcal{R}\right)_0=...=  \notag \\
&& =\left( (\lambda^{p+1}+\lambda^p u_0)dx\bigg| \lambda^{-2(p+1)} \left[%
\delta(x-s_1)\frac{\partial}{\partial x}, \delta(x-s_2)\frac{\partial}{%
\partial x} \right]_\mathcal{R} \right)_0=  \notag \\
&&=res_\lambda \int\limits_{0}^{2\pi}
(\lambda^{-p-1}+\lambda^{-p-2}u_0)(\delta(x-s_1)\delta^{\prime }(x-s_2)-
\delta(x-s_2)\delta^{\prime }(x-s_1))dx,  \notag
\end{eqnarray}
which is non vanishing only for the cases $p=-1$ and $p=0$. For the case 
$p=-1$ we have 
\begin{eqnarray}  \label{MP08}
&&\{ u_0(s_1),u_0(s_2) \}_{-1}=(\tilde{l}_{-1}\bigg|\lbrack \nabla
u_0(s_1),\nabla u_0(s_2) ]_\mathcal{R})_0=  \notag \\
&&=\left( \tilde{l}_{-1}\big| \left[\delta(x-s_1)\frac{\partial}{\partial x}%
, \delta(x-s_2)\frac{\partial}{\partial x}\right] _\mathcal{R}\right)_0= 
\notag \\
&&=\left( \tilde{l}_{-1}\bigg| \delta(x-s_1)\delta ^{\prime }(x-s_2)\frac{%
\partial}{\partial x}-\delta(x-s_2)\delta ^{\prime }(x-s_1)\frac{\partial}{%
\partial x} _\mathcal{R}\right)_0=  \notag \\
&&=res_\lambda \int\limits_{0}^{2\pi} \left< \left(1+\frac{u_0(x)}{\lambda}%
\right)dx \bigg| \delta(x-s_1)\delta ^{\prime }(x-s_2)\frac{\partial}{%
\partial x}-\delta(x-s_2)\delta ^{\prime }(x-s_1)\frac{\partial}{\partial x}%
\right>= \\
&&= \int\limits_{0}^{2\pi}u_0(x)(\delta(x-s_1)\delta ^{\prime
}(x-s_2)-\delta(x-s_2)\delta ^{\prime }(x-s_1))dx=  \notag \\
&&=u_0(s_1)\delta^{\prime }(s_1-s_2)-u_0(s_2)\delta ^{\prime
}(s_2-s_1)=u_0(s_1)\delta^{\prime }(s_1-s_2)+u_0(s_2)\delta ^{\prime
}(s_1-s_2)=  \notag \\
&&=\delta ^{\prime }(s_1-s_2)(u_0(s_1)-u_0(s_2)).  \notag
\end{eqnarray}
Taking into account, that $u_0=-2u_{s_1}$, we obtain, finally: 
\begin{eqnarray}  \label{MP09}
&&4\{ u_{s_1}(s_1),u_{s_2}(s_2) \}_{-1}=-2\delta ^{\prime
}(s_1-s_2)(u_{s_1}(s_1)+u_{s_2}(s_2)),  \notag \\
&&\{ u_{s_1}(s_1),u_{s_2}(s_2) \}_{-1}=-\frac 12\delta ^{\prime
}(s_1-s_2)(u_{s_1}(s_1)+u_{s_2}(s_2))=  \notag \\
&&=\frac 12\left( u_{s_2}(s_2)\theta^{\prime \prime }_{s_2s_1}(s_1-s_2) -
u_{s_1}(s_1)\theta^{\prime \prime }_{s_1s_2}(s_2-s_1)\right)\Rightarrow 
\notag \\
&&\{ u(s_1),u(s_2) \}_{-1}=\frac 12 \left( u_{s_2}(s_2)\theta(s_1-s_2) -
u_{s_1}(s_1)\theta(s_2-s_1)\right)
\end{eqnarray}
The same calculations one can perform for the case $p=0$: 
\begin{eqnarray}  \label{MP10}
&&\left\{ u_0(s_1),u_0(s_2)\right\}_0=\left( \tilde{l}_0\bigg| \lbrack
\nabla u_0(s_1),\nabla u_0(s_2) ]_\mathcal{R}\right)=  \notag \\
&&=\left( \tilde{l}_0 \bigg| \left[\lambda^{-1}\delta(x-s_1)\frac{\partial}{%
\partial x},\lambda^{-1}\delta(x-s_2)\frac{\partial}{\partial x} \right] _%
\mathcal{R}\right)=  \notag \\
&&=-\left( \tilde{l}_0 \bigg| \lambda^{-2} \left(\delta(x-s_1)\delta
^{\prime }(x-s_2)-\delta(x-s_2)\delta^{\prime }(x-s_1) \right) \right)= \\
&&=\int\limits_{0}^{2\pi} \left(\delta(x-s_1)\delta ^{\prime
}(x-s_2)-\delta(x-s_2)\delta^{\prime }(x-s_1) \right) dx=  \notag \\
&&=-\delta^{\prime }(s_1-s_2)+\delta^{\prime }(s_2-s_1)=-2\delta^{\prime
}(s_1-s_2),  \notag
\end{eqnarray}
and then finally 
\begin{eqnarray}  \label{MP11}
&& 4\left\{ u_{s_1}(s_1),u_{s_2}(s_2)\right\}_0=-2\delta^{\prime }(s_1-s_2) 
\notag \\
&& \left\{ u_{s_1}(s_1),u_{s_2}(s_2)\right\}_0=-\frac 12\delta^{\prime
}(s_1-s_2)  \notag \\
&& \left\{ u(s_1),u_{s_2}(s_2)\right\}_0=-\frac 12\delta(s_1-s_2)=\frac
12\theta^{\prime }_{s_2}(s_1-s_2) \Rightarrow  \notag \\
&& \left\{ u(s_1),u(s_2)\right\}_0=\frac 12\theta(s_1-s_2),
\end{eqnarray}
Having $\tilde{l}_0$ and reduced gradients \cite{HePrBlPr} 
\begin{align}  \label{MP12}
& \mathrm{grad\ } h^{(t)}(l)_{+}:=(\lambda^{2}\mathrm{grad\ } h)_{+}=\lambda^{2}+\lambda
u_{x}-u_{y}, \\
& \mathrm{grad\ } h^{(y)}(l)_{+}:=(\lambda^{1}\mathrm{grad\ } h)_{+}=\lambda+u_{x},  \notag
\end{align}
where 
\begin{equation}  \label{MP13}
\mathrm{grad\ } h(l)\sim1+u_{x}/\lambda-u_{y}/\lambda^{2}+O(1/\lambda^{3}),
\end{equation}
we can find the corresponding Hamiltonian equations: 
\begin{eqnarray}  \label{MP14}
\left\{ 
\begin{array}{l}
\tilde{l}_t=-ad_{ \mathrm{grad\ } h^{(t)}(l)_{+} }\tilde{l}_0 \\ 
\tilde{l}_y=-ad_{\mathrm{grad\ } h^{(y)}(l)_{+}}\tilde{l}_0%
\end{array}
\right. \Rightarrow \left\{ 
\begin{array}{l}
u_y=-\frac 32 u_x^2\equiv-\vartheta_0\mathrm{grad\ } H_0 \\ 
u_t=-\frac 52 u_x^3\equiv-\vartheta_{-1}\mathrm{grad\ }H_{-1}%
\end{array}
\right.
\end{eqnarray}

Using \eqref{MP14} and \eqref{MP10} we can find that 
\begin{eqnarray}  \label{MP15}
&&u_y=-\frac 32 u_x^2=-\frac 12
\int\limits_{-\infty}^{\infty}\theta(x-y)\mathrm{grad\ } H_0(y)dy= \\
&& = -\frac 12 \int\limits_x^\infty \mathrm{grad\ } H_0(x)dx=-\frac
12\partial^{-1}\mathrm{grad\ } H_0(x) \equiv -\vartheta_0\mathrm{grad\ } H_0,  \notag
\end{eqnarray}
which gives rise to the following expressions for $\vartheta_0$ and $\vartheta_0^{-1}$: 
\begin{equation}  \label{MP16}
\vartheta_0=\frac 12\partial^{-1}, \ \ \ \vartheta_0^{-1}=2\partial
\end{equation}
From \eqref{MP15} we have 
\begin{equation}  \label{MP17}
\mathrm{grad\ } H_0=\vartheta_0^{-1}\left(-\frac 32 u_x^2\right)=2\partial \left(-\frac
32 u_x^2\right)=-6 u_xu_{xx}
\end{equation}
Now we are able to restore the Hamiltonian $H_0$, using results of paper 
\cite{MPrF}: 
\begin{eqnarray}  \label{MP18}
H_0=-6 \int\limits_0^1
((\mu u)_x(\mu
u)_{xx}\big| u)d\mu 
\Rightarrow
 -\int\limits_{-\infty}^{\infty}(u_x^2)_x udx=
\int\limits_{-\infty}^{\infty}u_x^3dx  \notag
\end{eqnarray}

The similar calculations can be performed for Hamiltonian $H_{-1}$ and
Poisson operator $\vartheta_{-1}$: 
\begin{eqnarray}
&&u_t=-\frac 52 u_x^3=-\frac 12 \int\limits_{-\infty}^{\infty} \left( \frac{%
\partial u(y)}{\partial y}\theta(x-y)- \frac{\partial u(x)}{\partial x}%
\theta(y-x) \right) \mathrm{grad\ } H_{-1}(y)dy= \\
&& = 
-\frac 12 (\partial^{-1}u_x+u_x\partial^{-1})\mathrm{grad\ } H_{-1}(x) \equiv
-\vartheta_{-1}\mathrm{grad\ } H_{-1}  \notag
\end{eqnarray}
from which one can deduce, that 
\begin{eqnarray}
&&\vartheta_{-1}=\frac 12(\partial^{-1}u_x+u_x\partial^{-1}),\ \ \ \ \ \ \
\vartheta_{-1}^{-1}= 
\frac 12 \partial\frac{1}{\sqrt{u_x}}\partial^{-1}\frac{1}{\sqrt{u_x}}\partial
\end{eqnarray}
and 
\begin{eqnarray}
&&\mathrm{grad\ } H_{-1}=\vartheta_{-1}^{-1}\left(-\frac 52 u_x^3\right)= -6 u_x u_{xx}.
\end{eqnarray}
Finally, 
\begin{eqnarray}
H_{-1}=-6 \int\limits_0^1
((\mu u)_x(\mu
u)_{xx}\big| ud\mu) 
\Rightarrow
-\int\limits_{-\infty}^{\infty}(u_x^2)_xudx=
\int\limits_{-\infty}^{\infty}u_x^3dx\equiv H_0
\end{eqnarray}

\section{Plebański heavenly equation}

This equation \cite{Pleb} is 
\begin{equation}
u_{tx_{1}}+u_{yx_{2}}+u_{x_{1}x_{1}}u_{x_{2}x_{2}}-u_{x_{1}x_{2}}^{2}=0
\label{P1.29}
\end{equation}%
for a function $u\in C^{\infty }(\mathbb{R}^{2};\mathbb{T}^{2})$, where $%
(y,t;x_{1},x_{2})\in \mathbb{R}^{2}\times \mathbb{T}^{2}$. It was shown \cite%
{HePrBlPr} that the following \textquotedblright seed\textquotedblright
-element $\tilde{l}\in \mathcal{\tilde{G}}^{\ast }$, where $\mathcal{\tilde{G%
}}^{\ast }:=\widetilde{diff}^{\ast }(\mathbb{T}^{2})$, 
\begin{equation}
\tilde{l}=(\lambda +(u_{x_{1}}-u_{x_{2}})_{x_{1}})dx_{1}+(\lambda
+(u_{x_{1}}-u_{x_{2}})_{x_{1}})dx_{1},  \label{P1.30}
\end{equation}%
generates two independent Casimir functionals $h^{(1)},h^{(2)}\in I(\mathcal{%
\tilde{G}}^{\ast })$, whose gradients when $|\lambda |\rightarrow \infty $
are given by the expansions 
\begin{align}
& \mathrm{grad\ } h^{(1)}(\tilde{l})\sim
(0,1)^{T}+(u_{x_{2}x_{2}},-u_{x_{1}x_{2}})^{T}\lambda ^{-1}+O(\lambda ^{-2}),
\label{P1.31} \\
& \mathrm{grad\ } h^{(2)}(\tilde{l})\sim
(-1,0)^{T}+(-u_{x_{1}x_{2}},u_{x_{1}x_{1}})^{T}\lambda ^{-1}+O(\lambda
^{-2}).  \notag
\end{align}%
The Hamiltonian flows 
\begin{align}
& \tilde{l}_{t}=-ad_{\mathrm{grad\ } h^{(t)}(\tilde{l})}^{\ast }\tilde{l}
\label{P1.32} \\
& \tilde{l}_{y}=-ad_{\mathrm{grad\ } h^{(y)}(\tilde{l})}^{\ast }\tilde{l},
\end{align}%
where 
\begin{align}
& \mathrm{grad\ } h^{(t)}(\tilde{l}):=(\lambda \mathrm{grad\ }
h^{(1)}(l))_{+}=(u_{x_{2}x_{2}},\lambda -u_{x_{1}x_{2}})^{T},  \label{P1.33}
\\
& \mathrm{grad\ } h^{(y)}(\tilde{l}):=(\lambda \mathrm{grad\ } h^{(2)}(l))_{+}=(-\lambda
-u_{x_{1}x_{2}},u_{x_{1}x_{1}})^{T},  \notag
\end{align}%
and $\mathcal{\tilde{G}=\tilde{G}}_{+}\mathcal{\oplus \tilde{G}}_{-}$ with $%
\mathcal{\tilde{G}}_{-}|_{\lambda =\infty }=0$, are equivalent to the
following pair  of equations: 
\begin{align}
& (u_{x_{2}}-u_{x_{1}})_{t}=u_{x_{1}x_{1}}u_{x_{2}x_{2}}-u_{x_{1}x_{2}}^{2}
\label{P1.34} \\
& (u_{x_{1}}-u_{x_{2}})_{y}=u_{x_{1}x_{1}}u_{x_{2}x_{2}}-u_{x_{1}x_{2}}^{2},
\notag
\end{align}%
The compatibility condition of these flows \eqref{P1.33} is equivalent to
the following \cite{Pleb} vector field representation \cite{KruglMor} 
\begin{align}
& \frac{\partial \psi }{\partial t}+u_{x_{2}x_{2}}\frac{\partial \psi }{%
\partial x_{1}}+(\lambda -u_{x_{1}x_{2}})\frac{\partial \psi }{\partial x_{2}%
}=0,  \label{P1.35} \\
& \frac{\partial \psi }{\partial y}+(-\lambda -u_{x_{1}x_{2}})\frac{\partial
\psi }{\partial x_{1}}+u_{x_{1}x_{1}}\frac{\partial \psi }{\partial x_{2}}=0,
\notag
\end{align}%
satisfied for $\psi \in C^{2}(\mathbb{R}^{2}\times \mathbb{T}^{2};\mathbb{C}%
),$ any $(t,y;x_{1},x_{2})\in \mathbb{R}^{2}\times \mathbb{T}^{2}$ and all $%
\lambda \in \mathbb{C}$.

Let us take $\tilde{l}_{n}\in \mathrm{D}(\mathcal{\tilde{G}}^{\ast })$ 
\begin{equation}
\tilde{l}_{p}=\lambda ^{p}(\lambda +u_{0,x_{1}})dx_{1}+(\lambda
+u_{0,x_{1}})dx_{1}  \label{P1.36}
\end{equation}
with functional $u_{0}(s_{1},s_{2})=u_{s_{1}}-u_{s_{2}}$ as an independent
coordinate with parameters $s_{1}$ and $s_{2}$. It is easy to find, using
the similar notation as in \eqref{MP05}, that $p$ can be either $0$ or $-1$ and%
\begin{equation*}
\nabla u_{0}(s_{1},s_{2})=-\frac{1}{2\lambda }\left( \theta
(x_{1}-s_{1})\delta (x_{2}-s_{2})\frac{\partial }{\partial x_{1}}+\delta
(x_{1}-s_{1})\theta (x_{2}-s_{2})\frac{\partial }{\partial x_{1}}\right)
\end{equation*}

To find $\left\{ u_{0}(s_{1},s_{2}),u_{0}(s_{3},s_{4})\right\} _{0}$ one
will use the following chain of calculations:%
\begin{eqnarray}
&&\left\{ u_{0}(s_{1},s_{2}),u_{0}(s_{3},s_{4})\right\} _{0}=\left\{
u_{s_{1}}(s_{1},s_{2})-u_{s_{2}}(s_{1},s_{2}),u_{s_{3}}(s_{3},s_{4})-u_{s_{4}}(s_{3},s_{4})\right\} _{0}=
\\
&=&\left\{
u_{s_{1}}(s_{1},s_{2}),u_{s_{3}}(s_{3},s_{4})-u_{s_{4}}(s_{3},s_{4})\right\}
_{0}-\left\{
u_{s_{2}}(s_{1},s_{2}),u_{s_{3}}(s_{3},s_{4})-u_{s_{4}}(s_{3},s_{4})\right\}
_{0}= \nonumber\\
&=&\frac{\partial }{\partial s_{1}}\left\{
u(s_{1},s_{2}),u_{s_{3}}(s_{3},s_{4})-u_{s_{4}}(s_{3},s_{4})\right\} _{0}-%
\frac{\partial }{\partial s_{2}}\left\{
u(s_{1},s_{2}),u_{s_{3}}(s_{3},s_{4})-u_{s_{4}}(s_{3},s_{4})\right\}
_{0}\equiv  \nonumber\\
&\equiv &\frac{1}{2}\left( \delta (s_{1}-s_{3})\theta (s_{2}-s_{4})-\delta
(s_{2}-s_{4})\theta (s_{1}-s_{3})\right) = \nonumber\\
&=&\frac{1}{2}\frac{\partial }{\partial s_{1}}\left( \theta
(s_{1}-s_{3})\theta (s_{2}-s_{4})\right) -\frac{1}{2}\frac{\partial }{%
\partial s_{2}}\left( \theta (s_{1}-s_{3})\theta (s_{2}-s_{4})\right)
\Rightarrow  \nonumber\\
&&\left\{
u(s_{1},s_{2}),u_{s_{3}}(s_{3},s_{4})-u_{s_{4}}(s_{3},s_{4})\right\} _{0}=%
\frac{1}{2}\left( \theta (s_{1}-s_{3})\theta (s_{2}-s_{4})\right)
\Rightarrow  \\
&&(\partial /\partial s_{3}-\partial /\partial s_{4})\left\{
u(s_{1},s_{2}),u(s_{3},s_{4})\right\} _{0}=\frac{1}{2}\left( \theta
(s_{1}-s_{3})\theta (s_{2}-s_{4})\right) 
\end{eqnarray}%
and finally 
\begin{eqnarray*}
\left\{ u(s_{1},s_{2}),u(s_{3},s_{4})\right\} _{0} &=&\frac{1}{2}(\partial
/\partial s_{3}-\partial /\partial s_{4})^{-1}\left( \theta
(s_{1}-s_{3})\theta (s_{2}-s_{4})\right) = \\
&=&-\frac{1}{2}(\partial /\partial s_{1}-\partial /\partial
s_{2})^{-1}\left( \theta (s_{1}-s_{3})\theta (s_{2}-s_{4})\right) .
\end{eqnarray*}

For $n=-1$ we should find firstly the functionals $%
u(s_{1},s_{2}),u(s_{3},s_{4})$ and then obtain the corresponding explicit
expression for the Lie-Poisson bracket $\{u(s_{1},s_{2}),u(s_{3},s_{4})%
\}_{-1}$. From%
\begin{equation*}
\frac{\partial }{\partial s_{1}}\left( \frac{\partial u(s_{1},s_{2})}{%
\partial s_{1}}-\frac{\partial u(s_{1},s_{2})}{\partial s_{2}}\right)
=\left( \tilde{l}_{-1}|\frac{\partial }{\partial s_{1}}\theta
(s_{1}-x_{1})\delta (s_{2}-x_{2})\frac{\partial }{\partial x_{1}}+0\frac{%
\partial }{\partial x_{2}}\right) ,
\end{equation*}%
we immediately obtain  
\begin{equation*}
u_{0}(s_{1},s_{2}):=\frac{\partial u(s_{1},s_{2})}{\partial s_{1}}-\frac{%
\partial u(s_{1},s_{2})}{\partial s_{2}}=\left( \tilde{l}_{-1}\bigg|\theta
(s_{1}-x_{1})\delta (s_{2}-x_{2})\frac{\partial }{\partial x_{1}}\right) ,
\end{equation*}%
or 
\begin{equation}
\nabla u_{0}(s_{1},s_{2})(\tilde{l}_{-1})=\theta (s_{1}-x_{1})\delta
(s_{2}-x_{2})\frac{\partial }{\partial x_{1}},  \label{X1}
\end{equation}%
In the same manner we obtain the symmetric expresssion: 
\begin{equation*}
\frac{\partial u(s_{1},s_{2})}{\partial s_{1}}-\frac{\partial u(s_{1},s_{2})%
}{\partial s_{2}}=\left( \tilde{l}_{-1}\bigg|\delta (s_{1}-x_{1})\theta
(s_{2}-x_{2})\frac{\partial }{\partial x_{2}}\right) .
\end{equation*}%
We also have 
\begin{equation*}
\frac{\partial u(s_{3},s_{4})}{\partial s_{3}}-\frac{\partial u(s_{3},s_{4})%
}{\partial s_{4}}=\left( \tilde{l}_{-1}\bigg|\theta (s_{3}-x_{1})\delta
(s_{4}-x_{2})\frac{\partial }{\partial x_{1}}\right) ,
\end{equation*}%
and, respectively, 
\begin{equation}
\nabla u_{0}(s_{3},s_{4})(\tilde{l}_{-1})=\theta (s_{3}-x_{1})\delta
(s_{4}-x_{2})\frac{\partial }{\partial x_{1}},  \label{X2}
\end{equation}%
and also
\begin{equation*}
\nabla u_{0}(s_{3},s_{4})(\tilde{l}_{-1})=\delta (s_{3}-x_{1})\theta
(s_{4}-x_{2})\frac{\partial }{\partial x_{2}}.
\end{equation*}

For the Lie-Poisson bracket we can take\ (\ref{X1}) i (\ref{X2}):%
\begin{equation*}
\begin{array}{c}
\left\{ u_{0}(s_{1},s_{2}),u_{0}(s_{3},s_{4})\right\} =(\tilde{l}%
_{-1}|[\nabla u_{0}(s_{1},s_{2})(\tilde{l}_{-1}),\nabla u_{0}(s_{3},s_{4})(%
\tilde{l}_{-1})]_{R})= \\ 
=(\tilde{l}_{-1}|\theta (s_{1}-x_{1})\delta (s_{2}-x_{2})\frac{\partial }{%
\partial x_{1}}\left[ \theta (s_{3}-x_{1})\delta (s_{4}-x_{2})\right] \frac{%
\partial }{\partial x_{1}}- \\ 
-\theta (s_{3}-x_{1})\delta (s_{4}-x_{2})\frac{\partial }{\partial x_{1}}%
\left[ \theta (s_{1}-x_{1})\delta (s_{2}-x_{2})\right] \frac{\partial }{%
\partial x_{1}})= \\ 
=\ \left( \left[ 1+\lambda ^{-1}(u_{x_{1}}-u_{x_{2}})_{x_{1}}\right] dx_{1}+%
\left[ 1+\lambda ^{-1}(u_{x_{1}}-u_{x_{2}})_{x_{2}}\right] dx_{2}\right) \big|
\\ 
\big|-\theta (s_{1}-x_{1})\delta (s_{2}-x_{2})\delta (s_{3}-x_{1})\delta
(s_{4}-x_{2})\frac{\partial }{\partial x_{1}}+ \\ 
+\theta (s_{3}-x_{1})\delta (s_{4}-x_{2})\delta (s_{1}-x_{1})\delta
(s_{2}-x_{2})\frac{\partial }{\partial x_{1}})= \\ 
=\int \int dx_{1}dx_{2}\ [(u_{x_{1}}-u_{x_{2}})_{x_{1}}[\theta
(s_{3}-x_{1})\delta (s_{4}-x_{2})\delta (s_{1}-x_{1})\delta (s_{2}-x_{2})-
\\ 
-\theta (s_{1}-x_{1})\delta (s_{2}-x_{2})\delta (s_{3}-x_{1})\delta
(s_{4}-x_{2})]= \\ 
=(u_{s_{1}}-u_{s_{2}})_{s_{1}}\theta (s_{3}-s_{1})\delta
(s_{2}-s_{4})+(u_{s_{3}}-u_{s_{4}})_{s_{3}}\theta (s_{3}-s_{1})\ \delta
(s_{4}-s_{2})= \\ 
=[(u_{s_{1}}-u_{s_{2}})_{s_{1}}+(u_{s_{3}}-u_{s_{4}})_{s_{3}}]\theta
(s_{3}-s_{1})\ \delta (s_{4}-s_{2}).%
\end{array}%
\end{equation*}%
Finally,%
\begin{eqnarray}
\left\{ u_{0}(s_{1},s_{2}),u_{0}(s_{3},s_{4})\right\}  &=&\left( \frac{%
\partial }{\partial s_{1}}-\frac{\partial }{\partial s_{2}}\right) \left( 
\frac{\partial }{\partial s_{3}}-\frac{\partial }{\partial s_{4}}\right)
\left\{ u(s_{1},s_{2}),u(s_{3},s_{4})\right\} = \\
&=&[(u_{s_{1}}-u_{s_{2}})_{s_{1}}+(u_{s_{3}}-u_{s_{4}})_{s_{3}}]\theta
(s_{3}-s_{1})\ \delta (s_{4}-s_{2}), \nonumber
\end{eqnarray}%
or, equivavlently%
\begin{eqnarray}
\left\{ u(s_{1},s_{2}),u(s_{3},s_{4})\right\}  &=&\left( \frac{\partial }{%
\partial s_{1}}-\frac{\partial }{\partial s_{2}}\right) ^{-1}\left( \frac{%
\partial }{\partial s_{3}}-\frac{\partial }{\partial s_{4}}\right)
^{-1}\times  \\
&&\times \lbrack
(u_{s_{1}}-u_{s_{2}})_{s_{1}}+(u_{s_{3}}-u_{s_{4}})_{s_{3}}]\theta
(s_{3}-s_{1})\ \delta (s_{4}-s_{2}).  \nonumber
\end{eqnarray}

\section{Conclusion}

\bigskip 

\bigskip

\end{document}